\documentclass[12pt]{article}
\usepackage[a4paper,total={18cm,27cm}]{geometry}
\usepackage{graphicx}
\usepackage{authblk}
\usepackage{amsmath}
\usepackage[utf8]{inputenc}
\usepackage{hyperref}
\usepackage{verbatim}
\begin{document}


\title{
Detecting long-range interactions between migrating cells
}


\author[1]{C. Metzner}
\author[1]{F. Hörsch}
\author[1]{C. Mark}
\author[1]{T. Czerwinski}
\author[1]{A. Winterl}
\author[2,3,4]{C. Voskens}
\author[1]{B. Fabry}
\affil[1]{Biophysics, Friedrich-Alexander University Erlangen-Nürnberg, Erlangen}
\affil[2]{Department of Dermatology, University Hospital Erlangen, Friedrich-Alexander University Erlangen-Nürnberg, Erlangen}
\affil[3]{Comprehensive Cancer Center Erlangen - European Metropolitan Area of Nürnberg (CCC-ER-EMN), Erlangen, Germany}
\affil[4]{Deutsches Zentrum für Immuntherapie (DZI), Erlangen, Germany}

\maketitle


\begin{abstract}
Chemotaxis enables cells to systematically approach distant targets that emit a diffusible guiding substance. However, the visual observation of an encounter between a cell and a target does not necessarily indicate the presence of a chemotactic approach mechanism, as even a blindly migrating cell can come across a target by chance. To distinguish between the chemotactic approach and blind migration, we present an objective method that is based on the analysis of time-lapse recorded cell migration trajectories: For each movement step of a cell relative to the position of a potential target, we compute a $p$-value that quantifies the likelihood of the movement direction under the null-hypothesis of blind migration. The resulting distribution of $p$-values, pooled over all recorded cell trajectories, is then compared to an ensemble of reference distributions in which the positions of targets are randomized. First, we validate our method with simulated data, demonstrating that it reliably detects the presence or absence of remote cell-cell interactions. In a second step, we apply the method to data from three-dimensional collagen gels, interspersed with highly migratory natural killer (NK) cells that were derived from two different human donors. We find for one of the donors an attractive interaction between the NK cells, pointing to a cooperative behavior of these immune cells. When adding nearly stationary K562 tumor cells to the system, we find a repulsive interaction between K562 and NK cells for one of the donors. By contrast, we find attractive interactions between NK cells and an IL-15-secreting variant of K562 tumor cells. We therefore speculate that NK cells find wild-type tumor cells only by chance, but are programmed to leave a target quickly after a close encounter. We provide a freely available Python implementation of our p-value method that can serve as a general tool for detecting long-range interactions in collective systems of self-driven agents.
\end{abstract}

\vspace{1cm}
\noindent Correspondence to {\em claus.metzner@gmail.com}


\clearpage
\section*{Introduction}

Pursuit and evasion are ubiquitous in nature \cite{nahin2012chases}. An obvious example are predator-prey relations, where an agent attempts to catch a target that is struggling to escape. In such extreme cases, the continuous mutual reaction of the two opponents proves without doubt that they act in a goal-directed way and are able to sense each other from a distance. By contrast, when mobile agents are foraging for non-evading targets, it is not always clear whether the agents are performing a blind random walk and find their targets merely by chance, or if they recognize them already from larger distances and approach them systematically. The situation is particularly ambiguous in the case of micro-organisms, which can only migrate with a relatively large degree of directional randomness \cite{berg1993random}, and for which chemotaxis  is often the only available mechanism to locate distant targets \cite{eisenbach04}.

\vspace{0.5cm}
\noindent We have encountered such ambiguous behavior in experiments with highly mobile natural killer (NK) cells and almost immobile tumor cells, randomly distributed inside a 3-dimensional collagen gel (for details see below). By continuously monitoring the migration paths of the cells, we regularly observe NK cells that migrate to a nearby tumor cell, establish steric contact and attack the tumor cell, causing its subsequent death. This killing behavior after establishing cell-cell contact is consistent with the expected function of the immune system to eliminate pathogens. However, the interpretation of the preceding phase, in which the NK cells approach their targets, is ambiguous: On the one hand, it is known that NK cells can show a chemotactic response to suitable chemokine gradients \cite{taub1995alpha,loetscher1996activation}, and also that they can be chemotactically recruited by other cells of the immune system that have located a pathogen \cite{janeway1996immunobiology,kumar2011pathogen}. Given this fundamental chemotactic ability, the NK cells might very well be able to follow chemical traces of tumor cells directly. On the other hand, thorough visual inspection of time-lapse video recordings yields numerous examples where a migrating NK cell misses or even seems to turn away from a nearby tumor cell. 

\vspace{0.5cm}
\noindent The question of how immune cells are locating pathogens is of general importance, e.g. for optimizing cell-based immunotherapies for cancer \cite{schuster2006cancer,rosenberg2014decade,drake2014breathing}. A method for quantitatively analyzing the strength and range of chemotactic interaction could ultimately help opening ways to modify and improve the foraging efficiency of the immune cells by external interventions. 

\vspace{0.5cm}
\noindent In this study, we present a method for detecting long-range interactions between two types of agents, based only on their time-lapse recorded trajectories. The method could, in principle, be applied to any kind of self-propelled agents, including multi-cellular organisms or animals. However, we focus here on unicellular organisms, and specifically on immune cells and their targets. Our approach is as follows: We view the detection of interactions as a statistical test of the hypothesis that {\em immune cell migration is affected by distant targets}, against the null-hypothesis that the {\em immune cells perform a blind random walk}. Every discrete step of an immune cell's migration trajectory is characterized by a p-value, describing the probability that the step is part of a blind random walk. Computing the distribution of p-values over all recorded immune cells and time steps then reveals the presence of long-range interactions between immune cells and targets by a statistically significantly larger fraction of small p-values compared to a reference distribution of p-values in which the positions of the targets are randomized (Fig.~\ref{fig1}(c)). 

\vspace{0.5cm}
\noindent We validate our method with surrogate data, using a simulation framework for chemotactic behavior that has been published previously \cite{metzner2019efficiency}. In a first simulated scenario, called 'blind search' (BLS), the migration of immune cells is not influenced by the presence or the positions of targets. The immune cells migrate blindly with individually different but temporally constant migration parameters. In the second case, called 'random mode switching' (RMS), immune cells are also migrating blindly, but their migration behavior occasionally switches e.g. from slow to fast or from random to persistent, for reasons unrelated to the targets - a scenario that has been previously shown to be ubiquitous \cite{Metzner2015}. In the third test case, called 'temporal gradient search' (TGS), immune cells are able to detect differences of the target-related chemo-attractant concentration over time, and they modulate their degree of directional persistence accordingly - a well-known strategy of chemotaxis that is found in the run-and-tumble behavior of E.coli \cite{neidhardt1987escherichia}. In the fourth test case of 'spatial gradient search' (SGS), immune cells can directly measure and turn into the direction of a spatial gradient in the concentration of a chemo-attractant. Applied to these four simulated scenarios, the p-value method correctly finds the absence of long-range interactions between immune and target cells in the cases of 'blind search' and 'random mode switching', and the presence of interactions in the cases of 'temporal gradient search' and 'spatial gradient search'. 

\vspace{0.5cm}
\noindent In the next step, we turn to actual time-lapse recordings of human-derived natural killer (NK) cells, migrating within three-dimensional collagen gels (Fig.~\ref{fig1}(a,b)). A statistical analysis of the individual trajectories reveals that these cells are capable of migrating at large speeds exceeding 
10 $\mu$m/min \cite{zhou2017bystander}, and with various degrees of directional persistence, ranging from anti-persistent wiggling to almost uniform motion. When adding K562 tumor cells to the system, the NK cell's general migration behavior, described by the joint probability distribution of momentary speed and persistence, does not change significantly, even though we observe many events where NK cells encounter and kill some of the tumor cells. Finally, we apply our p-value method to test for potential chemotactic interactions. In the presence of wild-type K562 cells, which are known to have a rather low cytokine secretion, the NK cells do not show any evidence for attractive long-range interactions, but are, surprisingly, even repelled from the tumor cells in some cases. However, by replacing the tumor cells with a K562 variant that is secreting the chemokine IL-15 \cite{gong2010ex}, the NK cells (from one of the two human donors) show a strong chemotactic attraction. Morover, we also find attractive interactions among the NK cells themselves. The fact that each of these features is observed only in one of the donors points to a large degree of inter-personal variability in the behavior of NK cells.  


\section*{Materials and Methods}

\subsection*{Step 1: Data generation}

\subsubsection*{Experimental setup}

We assume an assay where both cell types are mixed together in a 3-dimensional matrix that enables effective cell migration and proper imaging with a microscope. If the matrix layer has a vertical thickness comparable to the expected 'killing radius' of the immune cells (the range around an immune cell in which all present target cells will be detected almost certainly), the search efficiency of the immune cells is limited by their ability to find the horizontal positions of the targets. The system can then be considered quasi two-dimensional, and the subsequent analysis can be restricted to the horizontal (x,y) cell positions. An additional advantage of this flat 3D geometry is that a possible durotactic response of the cells, driven by the stiffness gradient of the matrix in the z-direction, cannot interfere with chemotactic behavior: Even if one of the cell types preferentially migrates to the bottom and the other to the top of the gel, their vertical distance can never exceed the killing radius. In thicker matrices, a possible interference between durotactic and chemotactic effects can be avoided by excluding from the evaluation all cell detections in the top and bottom boundary layers of the matrix, where stiffness gradients can be present. A reasonable thickness of the excluded layers would then be $\approx 150\mu m$ in typical collagen gels. However, we have evaluated the data presented in this paper both with and without the boundary layers, and we obtained the same results in both cases. 

\vspace{0.5cm}
\noindent Note that the computation of the p-values is based on the x-y-coordinates of the immune cells only. The z-coordinates of the cells are required because we restrict the p-value analysis to immune-target pairs with a maximum 3D distance $r_{max}$. Repeating the analysis for different values of $r_{max}$ then allows us to estimate the actual radius of interaction. 

\vspace{0.5cm}
\noindent We assume that the microscope's field of view is time-lapse recorded with a constant time interval $\Delta t$ between successive frames. This $\Delta t$ has to be short enough ($<$1 min for fast cells), so that the cell configurations change only slightly from one frame to the next. Ideally, the outlines of the individual cells should still have some overlap in successive frames. Moreover, our method of interaction detection will work best with long overall recording periods ($\approx$300 frames), and with large numbers of cells in the field of view ($\approx$100). Missing frames are automatically handled (see below).

\vspace{0.5cm}
\noindent Our experiments are based on natural killer (NK) cells from human donors, which are in vitro activated and expanded \cite{mark2020cryopreservation}. A number of 300.000 NK immune cells and 120.000 K562 tumor cells are mixed with ice-cold 1500 $\mu$l acid-dissolved collagen solution (1.2mg/ml) and pipetted in each well of a tissue-culture treated 6-well plate (Corning). We used K562 wild-type tumor cells compared to K562-mbIL15-41BBL cells (gift from Prof. D. Campana, Department of Pediatrics, University Hospital Singapore; formerly St. Jude Children’s Research Hospital, Memphis, TN, USA). An increased amount of Interleukin 15 (IL 15) is found in the supernatant of K562-mbIL15-41BBL cells \cite{gong2010ex}, which is known to be important for NK cell activation, proliferation, survival and for an enhanced NK cell cyto-toxicity \cite{budagian200615, carson1994interleukin}. The polymerization of the collagen solution is initiated by placing the dish for 60 min in a cell culture incubator at 37$^\circ$C, 5$\%$ CO\textsubscript2. We perform z-scans (10 $\mu$m apart) through the $\approx$ 500 $\mu$m thick gel every 15 s for a duration of 30 min. Afterwards, another randomly chosen position is selected, and time-lapse imaging continues. In total, for each condition five positions are imaged. We recorded 2 independent data sets, each including 120 images, with a time interval of 15 seconds between subsequent frames. The images had 2752 x 2192 pixels with a size of 0.40954 $\mu$m. The procedure of cell extraction and all associated experimental protocols were approved by the ethics committee of the Friedrich-Alexander university Erlangen-Nürnberg (Project: 'Extracting rules of behavior in collective tumor cell systems', by Claus Metzner.). Informed consent was obtained from all participants. All methods were carried out in accordance with relevant guidelines and regulations.

\subsubsection*{Cell tracking}

To extract the information required for our interaction-detection algorithm, the recorded cells need to be detected and individually tracked, yielding the 3D center-of-mass coordinates $\vec{R}^{(i)}_t = (x^{(i)}_t,y^{(i)}_t,z^{(i)}_t)$ of every cell $i$ in each video frame $t$. Each cell needs to be labeled with a unique ID number $i$ that remains consistent over subsequent frames. Furthermore, each cell must be classified as either $c=0$ (immune cell) or $c=1$ (target cell). All information regarding a particular cell at a particular time needs to be stored in an 'observation' vector of the form $(t,x,y,z,i,c)$. The total number of observations may change between frames, as cells may leave or enter the microscope's field of view, because of cell division and death, or also due to tracking problems. The observations from all video frames (in any order) need to be combined into a matrix, with each row corresponding to an observation vector. This matrix, stored as a Numpy-array, forms the input to our interaction-detection algorithm. 

\vspace{0.5cm}
\noindent In this study, cells are segmented from the background and classified with a convolutional neural network, using the same method as described in \cite{mark2020cryopreservation}. The network has been trained on 6 manually labeled minimum/maximum intensity projection images of NK cells, mixed with K562 cells. After training, the network reached an object-wise F1-score of 0.84 on the training data. The center positions of the segmented cell areas in each frame were finally connected to trajectories, using Kalman filtering (Brownian motion model) for creating predictions and the Hungarian method to match the predictions with the detected center positions.

\subsubsection*{Generation of simulated data}

To validate our p-value method of interaction detection, we use a previously published software framework for the simulation of chemotactic hunting behavior \cite{metzner2019efficiency}, which provides the following four scenarios: (1) In 'Blind Search' (BLS), the immune cells do not interact at all with the targets but migrate blindly, according to a correlated random walk with fixed parameters for the mean step width (speed) and for the degree of directional persistence. (2) In 'Random Mode Switching' (RMS), the immune cells are still blind with respect to the targets, but occasionally switch between a highly persistent random walk and a non-persistent (diffusive) random walk mode. (3) In 'Temporal Gradient Sensing' (TGS), the immune cells actually approach the targets by following the temporal gradient of chemo-attractant. In particular, the model assumes that the immune cells perform a highly persistent correlated random walk as long as the concentration of chemo-attractant is increasing with time. When the concentration is decreasing, the immune cells switch to a diffusive (uncorrelated) random walk in order to find a more goal-directed migration direction. (4) In 'Spatial Gradient Sensing' (SGS), the immune cells are able to measure the spatial gradient of chemo-attractant and to actively turn into the direction of a nearby target. 

\vspace{0.5cm}
\noindent To generate the surrogate data for the present paper, we set all parameters of the chemotactic simulating framework to the same values as in \cite{metzner2019efficiency}. However, while all results were averaged over 10000 runs in \cite{metzner2019efficiency}, we now produce for each scenario only a single simulated data set with a longer duration (500 time steps of $\Delta t=1$min), with a larger field of view (5000 $\mu$m x 5000 $\mu$m) and with a larger number of cells (100 immune cells and, initially, 50 target cells).

\subsection*{Step 2: Data filtering}

\subsubsection*{'Triplet'-based analysis}

The elementary unit for our data analysis is a 'triplet', consisting of three consecutive cell positions 
$\left\{\;\vec{R}^{(i)}_{t\!-\!1}, \vec{R}^{(i)}_t, \vec{R}^{(i)}_{t\!+\!1}\;\right\}$
of an immune cell $i$ (solid blue dots in Fig.~\ref{fig1}(d)). Our algorithm automatically extracts all such triplets from the matrix of observations. Recording or tracking gaps are automatically excluded from the list of triplets.

\subsubsection*{Excluding cells with too few triplets}

From the list of all triplets, we extract the subset belonging to a particular immune cell $i$. If this cell has a number $N^{(tr)}_i$ of triplets smaller than $N^{(tr)}_{min}$ (typical settings range from 5 to 20), the cell is excluded from the subsequent analysis, because it is not possible to reliably estimate the average migration behavior of a cell based on such a small number of positions (for details see below).

\subsubsection*{Excluding triplets with too distant targets}

Often, there will be a-priori knowledge about the maximum expected distance $r_{max}$ for interactions between immune and target cells (Otherwise, $r_{max}$ can be set to a value larger than the size of the microscope's field of view). For the purpose of interaction detection (for details see below), we restrict the analysis to triplets which have at least one target present in a sphere of radius $r_{max}$ around the triplet's central immune cell position $\vec{R}^{(i)}_t = (x^{(i)}_t,y^{(i)}_t,z^{(i)}_t)$. Generally, a smaller $r_{max}$ reduces the computation time of the algorithm, but a larger $r_{max}$ increases the number of possible targets and thus reduces statistical fluctuations in the evaluation.

\subsection*{Step 3: Data analysis}

\subsubsection*{Cell migration model}

We focus on the in-plane, horizontal motion of the cells. For this purpose, we only use two-dimensional coordinates, in the following denoted by lower-case position vectors $\vec{r}^{(i)}_t = (x^{(i)}_t,y^{(i)}_t)$. The sequence of horizontal positions  $\vec{r}$ of each individual cell $i$ over successive time indices $t=0,1,2,\ldots$ is approximated by a directionally persistent, discrete time random walk. It is characterized by a certain distribution $p_i(w)$ of step widths $w$, and a distribution $p_i(\theta)$ of turning angles $\theta$. Here, the step width (Euclidean distance) for a cell's movement between time $t$ and $t\!+\!1$ is defined as $w=|\vec{r}^{(i)}_{t\!+\!1}-\vec{r}^{(i)}_t|$, and the turning angle is defined as the angle between the two shift vectors $\left[\vec{r}^{(i)}_{t\!+\!1}-\vec{r}^{(i)}_t\right]$ and $\left[\vec{r}^{(i)}_t-\vec{r}^{(i)}_{t\!-\!1}\right]$.

\noindent The step width distribution is modeled as a Rayleigh distribution with speed parameter $\sigma_i$:
\begin{equation}
p_i(w) = \frac{w}{\sigma_i^2} \exp\left( -\frac{1}{2}\frac{w^2}{\sigma_i^2}\right).
\end{equation}

\noindent The turning angle distribution is modeled as a von-Mises distribution with persistence parameter $\kappa_i$:
\begin{equation}
p_i(\theta) = \frac{1}{2\pi I_0(\kappa_i)} \exp\left( \kappa_i\cdot\cos(\theta) \right).
\end{equation}

\noindent We estimate the individual speed parameter $\sigma_i$ and persistence parameter $\kappa_i$ for each cell $i$, based on its complete recorded time series, as described in \cite{evans2000statistical}. These two parameters characterize the average in-plane migration behavior of the cell.

\subsubsection*{'Ordinary' and 'extraordinary' steps}

In our method of interaction detection, we focus on the turning angles $\theta$ of the immune cells, a quantity that is statistically fluctuating from one step to the next, approximately described by the von-Mises distribution with persistence parameter $\kappa$. For positive $\kappa$, corresponding to directionally persistent migration, the von-Mises distribution is peaked around a zero turning angle, so that most of the turning angles will have small magnitudes ('ordinary moves') and only relatively few will have large magnitudes  ('extraordinary moves').

\vspace{0.5cm}
\noindent For the sole purpose of visualization, we can set an arbitrary threshold angle $\theta_{thr}$ and define ordinary moves as those with  $\left| \theta \right| \leq \theta_{thr}$. For example, the threshold $\theta_{thr}$ could be chosen such that ordinary moves occur in target-blind migration with a probability of $P_{ord}=0.95$. Graphically, the interval of ordinary turning angles can then be depicted as a 'persistence cone' (Fig.~\ref{fig1}(d,e)). Turning angles that lead to the outside of the persistence cone would then be regarded as extraordinary (Case (3) in Fig.~\ref{fig1}(e)). Note, however, that our method of interaction detection is directly based on the von-Mises distribution, and neither $\theta_{thr}$ nor $P_{ord}$ play any role in the calculation of the p-values. The definition of the persistence cone is only used to illustrate the fundamental idea of the method.

\subsubsection*{Evidence for target pursuit}

\begin{figure}[h!]
\centering
\includegraphics[width=14cm]{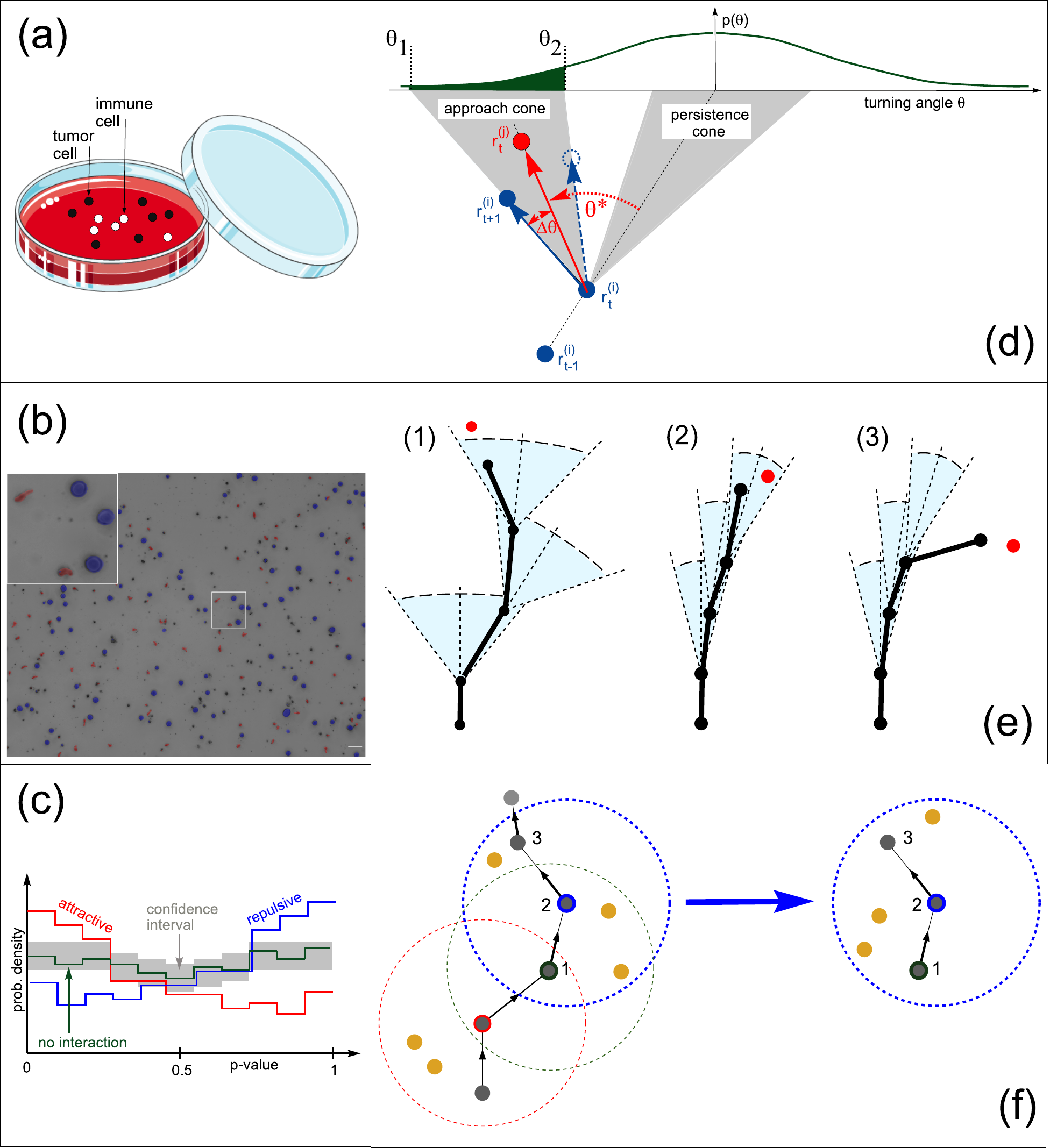}
\caption{
{\bf (a)} Tumor cells and immune cells are randomly dispersed within a suitable 3D matrix. {\bf (b)} Cells are automatically detected, classified and tracked over time. In the example frame, K562 tumor cells are labeled in blue, NK cells in red, non-cell objects in black. The white scale bar in the lower right corner corresponds to 50 $\mu$m. The inset shows the different cell morphologies. {\bf (c)} A p-value distribution is computed from the data and compared with a confidence interval corresponding to randomized data. The sketch shows the characteristic signatures of attractive (red) ane repulsive (blue) interactions. {\bf (d)} Computation of the p-value (green shaded area under the curve $p(\theta)$). We consider a 'triplet', consisting of three consecutive positions ($\vec{r}_{t\!-\!1}^{(i)}$, $\vec{r}_{t}^{(i)}$, $\vec{r}_{t\!+\!1}^{(i)}$) of the focal immune cell (blue solid circles) and the position $\vec{r}_t^{(i)}$ of a target cell (red circle) in the vicinity. The 'persistence cone' (right shaded area) is the interval of the most probable migration directions of a target-blind immune cell, which is determined by the previous migration direction (between time step $t\!-\!1$ and $t$), and by the known turning angle distribution $p(\theta)$ (olive curve on the top). The quantity $\theta^{\star}$ (red) is the optimal turning angle that would align the immune cell precisely towards the target cell. The approach cone (left shaded area) is the interval of migration directions which are at least as target-oriented as the actual move of the immune cell. By integrating $p(\theta)$ over the approach cone (that is, from turning angle $\theta_1=\theta^{\star}-\Delta\theta$ to $\theta_2=\theta^{\star}+\Delta\theta$), we compute a p-value, {\em the probability that a move at least as target-directed as observed could occur in a target-blind random walk}.
{\bf (e)} Three examples of immune cell trajectories (black) in relation to a target cell (red). Cases (1) and (2) are not indicative of goal-directed migration, but case (3) is 'suspicious', because the cell steps out of its persistence cone and is at the same time very target-directed.
{\bf (f)} A reference system without interactions between immune and target cells is generated by re-positioning the target cells randomly, while leaving the immune cell trajectory unchanged.
\label{fig1}}
\end{figure}

Even if an immune cell is moving with high persistence into the direction of a nearby target (red dots in Fig.~\ref{fig1}(e)), this provides no evidence for target pursuit if the individual migration steps are classified as ordinary (cases (1) and (2) in Fig.~\ref{fig1}(e)). Only steps that are classified as extraordinary and at the same time are highly target-directed provide some evidence for target pursuit (case (3) in Fig.~\ref{fig1}(e)), in particular if they occur more often than would be expected for a target-blind immune cell.

\subsubsection*{Definition of the 'approach cone'}

To quantify the target-directedness of a step, we define an 'approach cone' (left gray area in Fig.~\ref{fig1}(d)) as follows: We consider three successive positions of an immune cell $i$ (solid blue circles), given by the triplet $\left\{\;\vec{r}^{(i)}_{t\!-\!1}, \vec{r}^{(i)}_t, \vec{r}^{(i)}_{t\!+\!1}\;\right\}$, as well as the position $\vec{r}^{(j)}_t$ of a nearby target cell $j$ (solid red circle). The optimal turning angle $\theta^{\star}$ (dashed red arc) of a goal-directed immune cell shifts the migration trajectory directly towards the target cell with an optimal shift vector $\vec{s}_{opt}\!=\!\vec{r}^{(j)}_t\!-\!\vec{r}^{(i)}_t$ (red vector). In practice, immune cell $i$ has moved along the vector $\vec{s}_{1}\!=\!\vec{r}^{(i)}_{t\!+\!1}\!-\!\vec{r}^{(i)}_t$ (solid blue vector), which encloses an angle $\Delta\theta$ (red arc) with the optimal shift vector $\vec{s}_{opt}$. There exists another, hypothetical shift vector $\vec{s}_2$ that encloses the same angle $\Delta\theta$ with the optimal shift vector $\vec{s}_{opt}$ (dashed blue vector). The interval of turning angles between $\theta_1\!=\!\theta^{\star}\!-\!\Delta\theta$ and $\theta_2\!=\!\theta^{\star}\!+\!\Delta\theta$ ia the approach cone, that is, the set of directions which are at least as target-oriented as the actual shift of the immune cell.

\subsubsection*{Definition and interpretation of the p-value}

By integrating the von-Mises distribution $p_i(\theta)$ over all turning angles $\theta\in\left[\;
\theta^{\star}\!-\!\Delta\theta,\; \theta^{\star}\!+\!\Delta\theta\;
\right]$ within the approach cone, we compute a p-value, subsequently denoted by the symbol $\hat{p}$ (green area under the $p(\theta)$ curve in Fig.~\ref{fig1}(d)). 
\begin{equation}
\hat{p} = \int_{\theta^{\star}\!-\!\Delta\theta}^{\theta^{\star}\!+\!\Delta\theta} p_i(\theta) d\theta.
\end{equation}
$\hat{p}$ can be interpreted as {\em the probability that the observed move of the immune cell, or an even more target-directed move, could occur in a target-blind migration}. Very small p-values indicate that immune cells are attracted towards target cells, while very large p-values indicate that immune cells are repelled from target cells. Due to its definition, the p-value can be very small only if three conditions are simultaneously fulfilled: (A) The persistence cone is narrow (high directional persistence of the immune cell, corresponding to a narrow von-Mises distribution $p_i(\theta)$). (B) The approach cone is narrow (the immune cell turns almost exactly towards the target cell). (C) The two cones are non-overlapping and distant from each other (the immune cell is 'going out of its way' to approach the target). 

\subsubsection*{Distribution of p-values}

Finding just a few steps with very low p-value does in general not provide convincing evidence for a target-directed immune cell migration. Moreover, a subset of immune cells might be attracted to the targets while others are repelled from them. Alternatively, the same immune cell could be attracted and repelled by targets at different times. All these cases are comprehensively described by the global probability distribution $q_{obs}(\hat{p})$ of all observed p-values. We approximate this continuous probability distribution by a discrete histogram (see, fir example, Fig.~\ref{fig2}). 

\subsubsection*{Reference distributions of p-values}

Since extraordinary steps occur also in target-blind migration with probability $1-P_{ord}$, and some some of these steps may accidentally lead into the direction of nearby targets, we need to compare $q_{obs}(\hat{p})$ with a reference distribution $q_{ref}(\hat{p})$ of a system that resembles the observed one in all respects, except that there are no interactions between immune cells and target cells. To obtain this reference distribution, we use the following bootstrapping method \cite{westfall1993resampling}: We generate a reference data set by keeping the positions of the immune cells unchanged but shifting all target cells that are located within the maximum interaction radius $r_{max}$ to new, independent random positions within that radius (Fig.~\ref{fig1}(f). From this reference data set, we compute the histogram of p-values, yielding a (first) reference distribution $q_{ref}(\hat{p})$ with the same sample size as $q_{obs}(\hat{p})$. Since we are interested in the fluctuations of the $q_{ref}(\hat{p})$-values in each histogram bin, we repeat the same procedure for a large number $N_s \approx 100$ of statistically independent reference data sets. From these $N_s$ histograms, we compute the mean $\mu_k$ and standard deviation $\sigma_k$ of the $q_{ref}(\hat{p})$-values in each histogram bin $k$. Based on this statistics, we define confidence intervals (gray shaded area in Fig.~\ref{fig1}(c) and in all subsequent p-value distributions) for each bin $k$ as $\left[\;\mu_k-1.645\;\sigma_k,\;\mu_k+1.645\;\sigma_k\;\right]$. Assuming a normal distribution, the probability of a value above the upper (or below the lower) limit of the confidence interval is then 0.05 in the target-randomized reference system. If the measured p-value distribution $q_{obs}(\hat{p})$ lies outside the confidence interval of the reference systems at least in some histogram bins, this may be interpreted as a statistically significant effect, indicating that the targets somehow affect the migration of the immune cells. 

\vspace{0.2cm}
\noindent In our case, the width $\Delta q_k$ of the confidence intervals has been arbitrarily set to $2\times1.645\;\sigma_k$, because the resulting significance level of 0.05 is a common choice in science. The user is however free to choose other values for $\Delta q_k$, such as $\Delta q_k = 2\times2.325\;\sigma_k$ for a significance level of 0.01.


\section*{Results}

\subsection*{Validation of p-value method with simulated data}

\begin{figure}[h!]
\centering
\includegraphics[width=17cm]{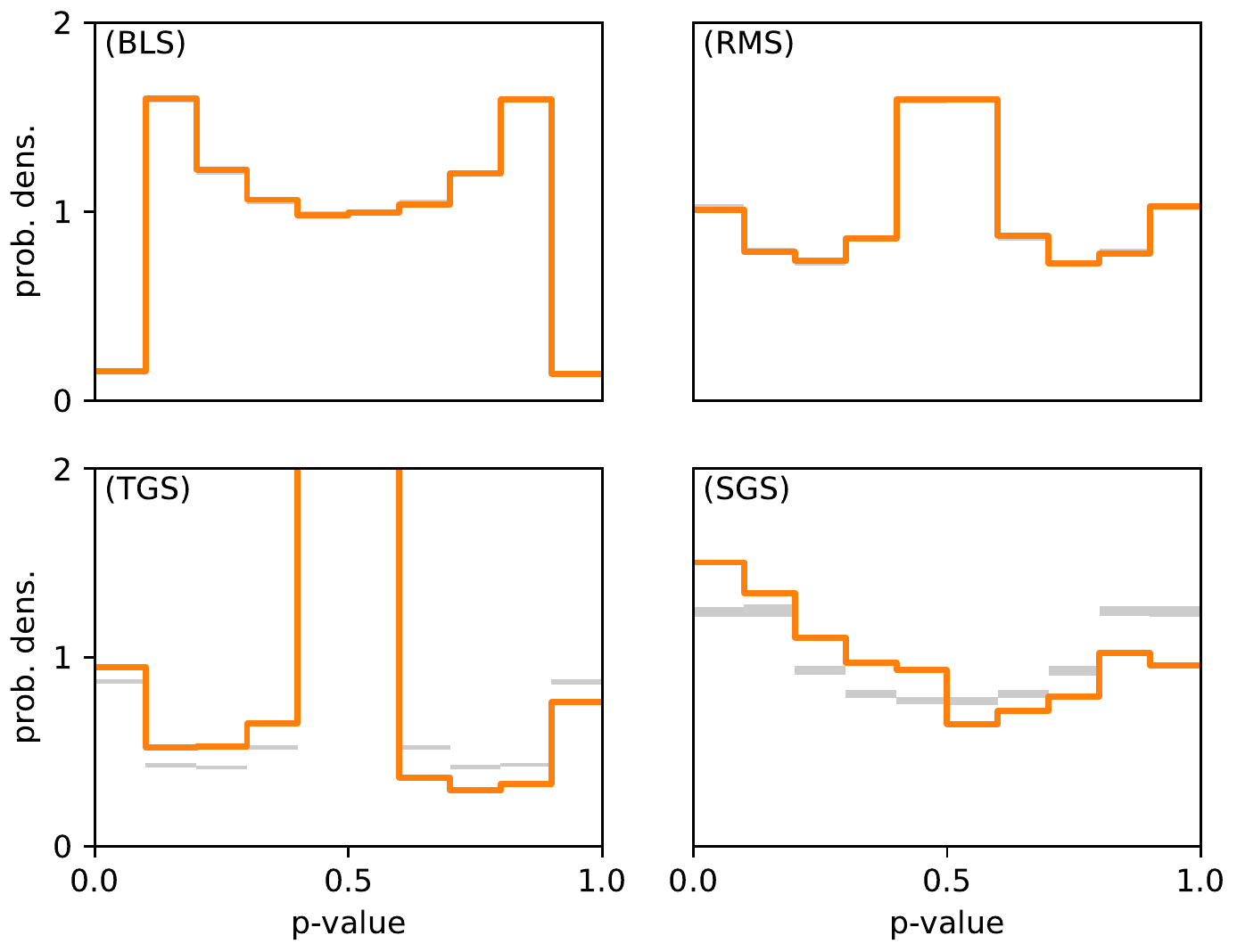}
\caption{
Application of the p-value method to four different types of simulated data. Shown are in each case the p-value distributions $q_{obs}(\hat{p})$ of the actual system (orange lines) and the confidence intervals (gray shaded areas), based on the distributions $q_{ref}(\hat{p})$ of the randomized reference systems. Values above (or below) the confidence interval occur with a probability of 5 percent in the randomized system. 
{\bf (a)} Blind search (BLS): Simulated immune cells migrate blindly, according to a correlated random walk with temporally constant migration parameters.
{\bf (b)} Random mode switching (RMS): Simulated immune cells migrate blindly, according to a correlated random walk with temporally fluctuating migration parameters.
{\bf (c)} Temporal gradient sensing (TGS): Simulated immune cells use temporal gradients of a chemo-attractant to pursue the target cells.
{\bf (d)} Spatial gradient sensing (SGS): Simulated immune cells use spatial gradients of a chemo-attractant to pursue the target cells.
In cases (a) and (b), where there are no interactions between simulated immune and target cells, the actual p-value distributions are inside the confidence intervals of the randomized system. In cases (c) and (d), chemotactic interactions between simulated immune and target cells lead to significant differences between the observed and reference distributions.
\label{fig2}}
\end{figure}

We first validate the p-value method with data from chemotaxis simulations, using a recording period of $\Delta t=1$ min, a maximum detection radius of $r_{max}=500\;\mu$m, and a number of $N_s=100$ reference distributions. In the BLS scenario, as expected, the resulting p-value distribution is almost identical to that of the randomized reference system (Fig.~\ref{fig2}(a)). 
In the RMS scenario, the overall shape of the p-value distribution is different from the BLS case, because RMS is a heterogeneous random walk. Nevertheless, the observed and reference distributions are again almost identical (Fig.~\ref{fig2}(b)).
In the TGS scenario, the shape of the p-value distribution is similar to that of RMS, because both search strategies share the feature of mode switching, one being controlled by chemoattractant gradients, the other occurring just randomly. Now, however, there are significant differences between the observed and reference distributions (Fig.~\ref{fig2}(c)). In particular, the observed distribution shows a larger probability of p-values smaller than 1/2, thus indicating attractive interactions. 
In the SGS scenario, we find yet another shape of the p-value distribution, but again
there are significant differences between the observed and reference distributions (Fig.~\ref{fig2}(d)).

\subsubsection*{Effect of recording interval $\Delta t$
on p-value distributions}

Even if the motion of cells appears non-directional on short time scales, a target-directed migration may nevertheless emerge on larger time scales. To test for this possibility, we can sub-sample the recorded data by evaluating the triplets at time points $t-n\Delta t$, $t$, and $t+n\Delta t$, with an integer number $n$, thus effectively increasing the recording time period to $n\Delta t$. When applying this sub-sampling approach to the surrogate data simulated in the SGS scenario (columns of Fig.~\ref{fig3}), we indeed find that the differences between the observed versus reference distributions become more pronounced for larger effective recording intervals.

\subsubsection*{Effect of maximum detection radius $r_{max}$
on p-value distributions}

\begin{figure}[h!]
\centering
\includegraphics[width=17cm]{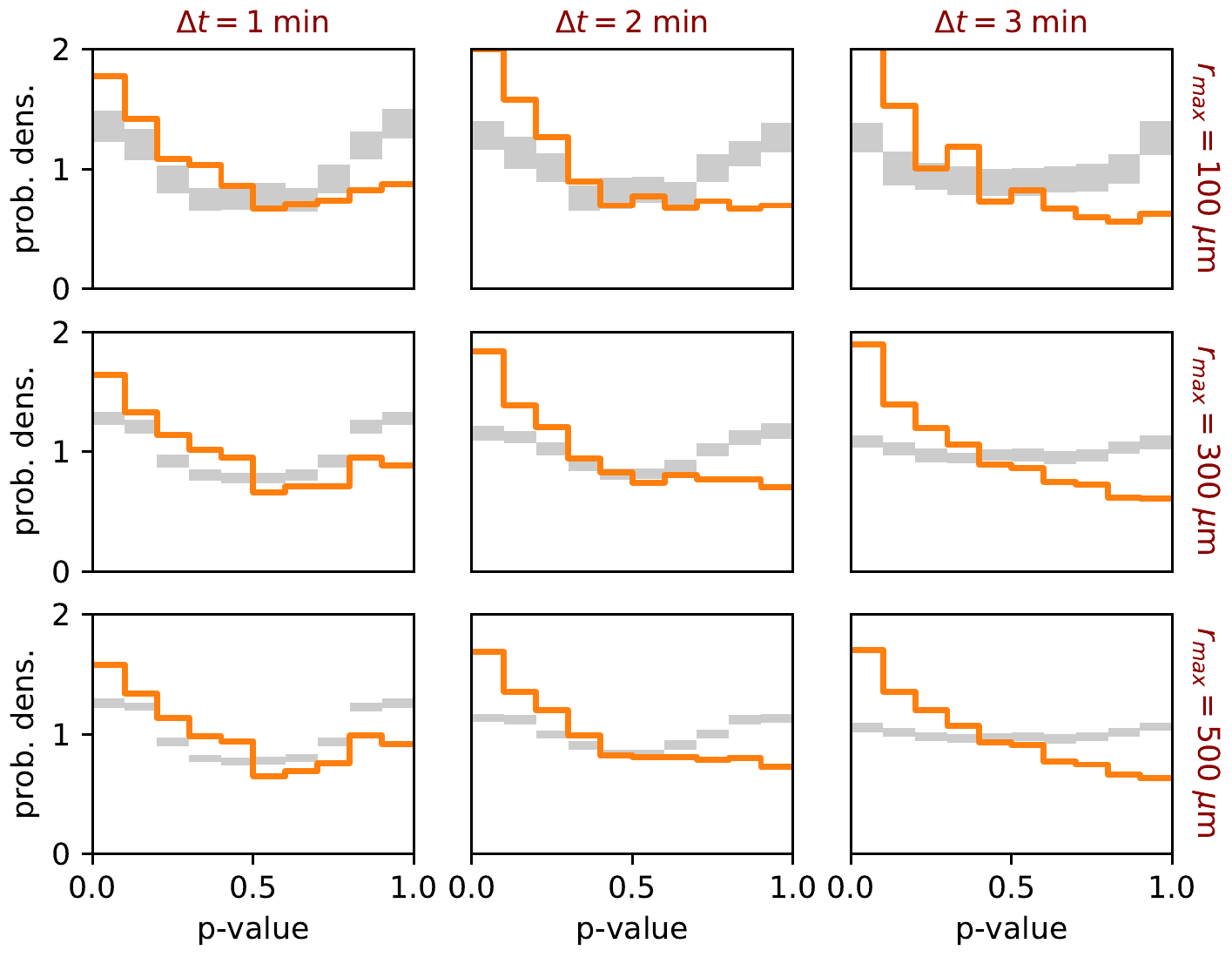}
\caption{
Application of the p-value method to data simulated in the SGS model, using different recording intervals $\Delta t$ (columns) and maximum interaction distances $r_{max}$ (rows). The differences between the simulated and reference data are more pronounced for larger recording intervals. Increasing the maximum interaction distance helps to reduce the width of the confidence interval in the reference distributions.
\label{fig3}}
\end{figure}

We also test the effect of the maximum detection radius $r_{max}$
on the p-value distributions for the SGS simulations (rows of Fig.~\ref{fig3}). The results demonstrate that a larger $r_{max}$ is generally preferable, because it reduces the widths of the confidence intervals. When applying our method to new systems in which the range of interactions is unknown, we therefore recommend to set $r_{max}$ to the diagonal size of the field of view.

\subsubsection*{Signature of weakly repulsive and weakly attractive interactions}

\begin{figure}[h!]
\centering
\includegraphics[width=17cm]{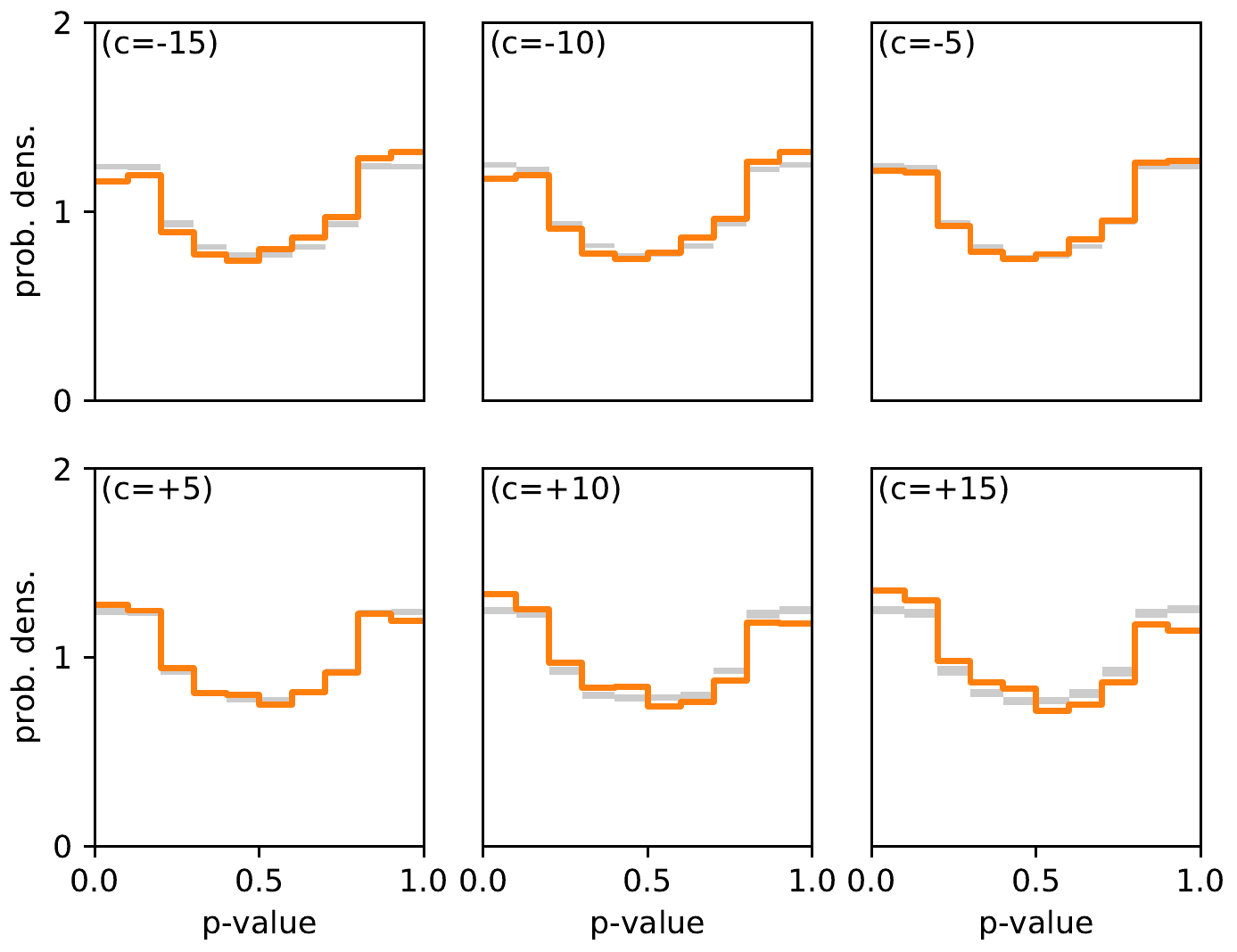}
\caption{
Application of the p-value method to simulated data (modified SGS model), where  tumor cells are for the immune cells weakly repulsive (top row, negative values of the chemotaxis response coefficient $c$), or weakly attractive (bottom row, positive values of the chemotaxis response coefficient $c$). Note that in the standard SGS model (Fig.~\ref{fig2}(d)), the coefficient is $c=+$500. For repulsive interactions, there are fewer small p-values and more large p-values than in the reference systems. 
\label{fig4}}
\end{figure}

Next we consider the case of very weak interactions between immune and target cells, using again the simulated data in the SGS scenario. In order to modulate the interaction strength, we vary the chemotactic response parameter $c$ (denoted by $c_R$ in \cite{metzner2019efficiency}), which controls how sensitively the immune cells turn into the direction of the chemotactic concentration gradient. We find that for attractive interactions (positive $c$, bottom row of Fig.~\ref{fig4}), p-values smaller than 1/2 are still more frequent than in the reference system, but the differences eventually become non-significant in the case of very weak attraction (case $c=+5$ in Fig.~\ref{fig4}). Conversely, in the case of repulsive interactions (negative $c$, top row of Fig.~\ref{fig4}), p-values smaller than 1/2 are less frequent than in the reference system.

\subsection*{Experiments with immune and tumor cells in thick collagen}

As a first practical test, we apply our method to time-lapse recordings of cells that are randomly dispersed within collagen gels. In the main part of the paper, we focus on thick, genuinely three-dimensional gels (height $\approx$ 500 $\mu$m), but in the Supplemental Information we also present data measured in flat, effectively two-dimensional gels (height $\approx$ 30 $\mu$m). Within the thick gels, we compare the behavior of the immune cells in three different systems: (1) NK cells only, (2) NK cells mixed with wild type K562 cells, and (3) NK cells mixed with K562-mbIL15-41BBL cells that secrete an increased amount of Interleukin 15 (IL 15). Except for the tumor cell type, all conditions, in particular the cell densities, are identical in the three systems. These experiments are performed twice, using NK cells from two different human donors.

\subsubsection*{General NK cell migration parameters}

\begin{figure}[h!]
\centering
\includegraphics[width=17cm]{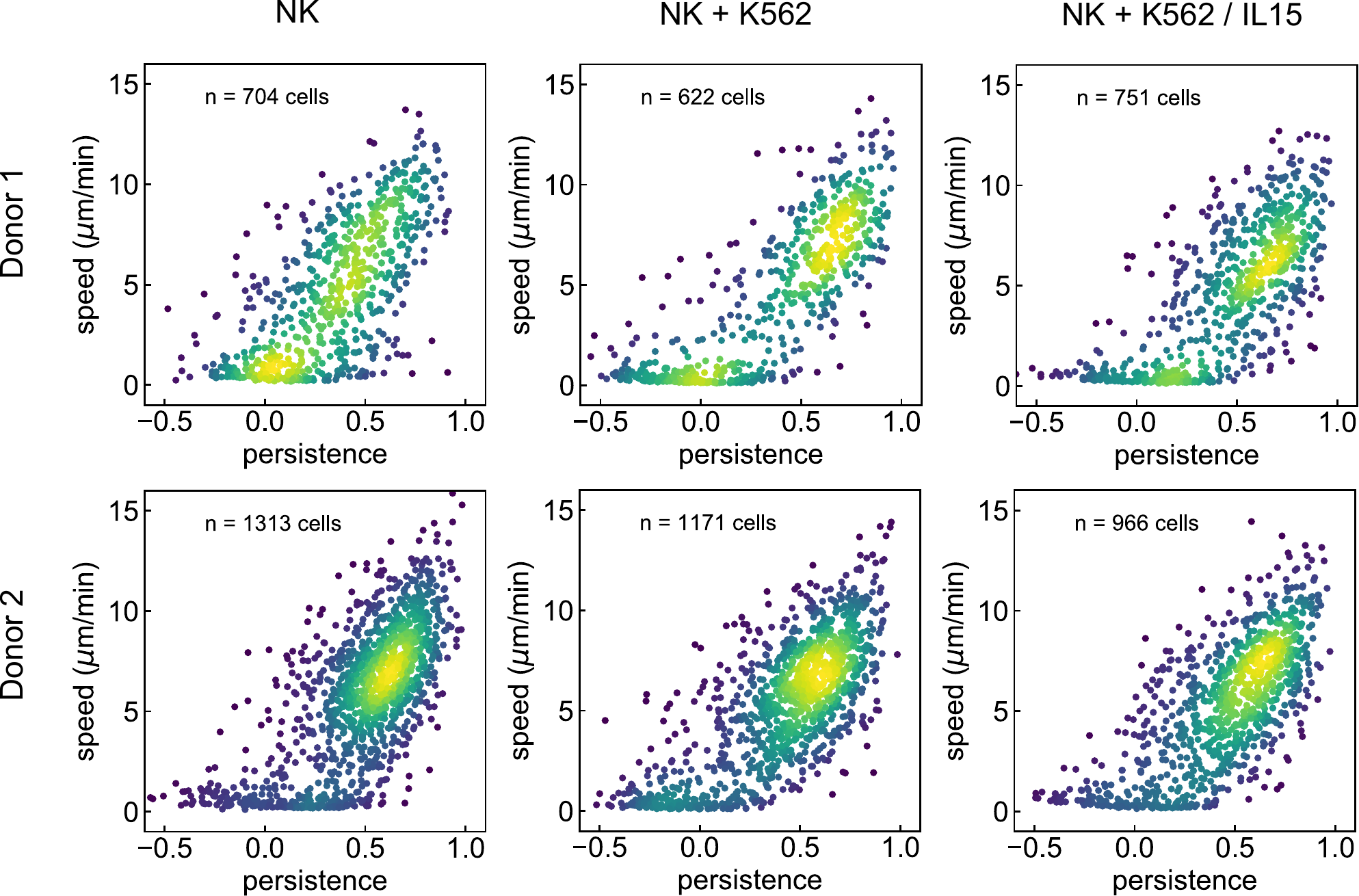}
\caption{Momentary migration parameters of NK cells in a thick collagen gel. Shown are the joint probability distributions $p(\cos(\theta),v)$ of persistence (approximated by the cosine of the turning angle) and speed for a system containing only NK cells (left column), for a system of NK cells mixed with wild-type K562 cells (center column), and for a system of NK cells mixed with K562 cells that secret the chemokine IL-15 (right column). Identical experiments have been performed using NK cells from two different donors (rows). The colors represent a kernel density estimation of the continuous probability density. There are no significant differences between the distributions $p(\cos(\theta),v)$ in the six different cases.
\label{fig5}}
\end{figure}

First, we investigate the statistical distribution of the NK cell's momentary migration properties, characterized by the speed $v$ (defined as the step width $w$ from one frame to the next, divided by the recording time interval $\Delta t$) and by the cosine of the turning angle $\theta$. The latter quantity is a simple measure of directional persistence, which can range from $-1$ (perfectly anti-directional back-and-forth motion), over $0$ (non-directional motion), to $+1$ (perfectly directional, straight-forward motion). After computing the pairs 
$\left(\cos(\theta^{(i)}_t),v^{(i)}_t\right)$ 
of momentary migration parameters for all immune cells $i$ and time steps $t$ (excluding very short trajectories from cells that could not be tracked for at least 20 consecutive steps), their joint probability distribution $p(\cos(\theta),v)$ is estimated using kernel density approximation. We perform the identical evaluation for the three types of systems (NK, NK-K562 and NK-K652/IL15; columns in Fig.~\ref{fig5}) and for the two human donors (rows in Fig.~\ref{fig5}). The resulting distributions can be divided into two distinct fractions: one consisting of immobile NK cells with speeds close to zero and persistences roughly in the range [-0.5,0.5], and another fraction consisting of mobile cells. For the latter, speed and persistence are positively correlated, with a few cells reaching momentary speeds $v$ well above 10 $\mu$m/min and persistences close to one. We find basically the same $p(\cos(\theta),v)$ for all three types of systems, demonstrating that possible long-range interactions between NK and tumor cells do not significantly affect the joint distribution of the migration parameters. Indeed, the joint distribution only describes the fundamental spectrum of possible movement steps that are available for a given cell type. Chemotaxis is a higher-order correlation effect, which comes about by choosing, out of this set, an optimal temporal sequence of movement steps that lead towards (or away from) a target.   

\subsubsection*{p-value distributions of NK cells}

\begin{figure}[h!]
\centering
\includegraphics[width=17cm]{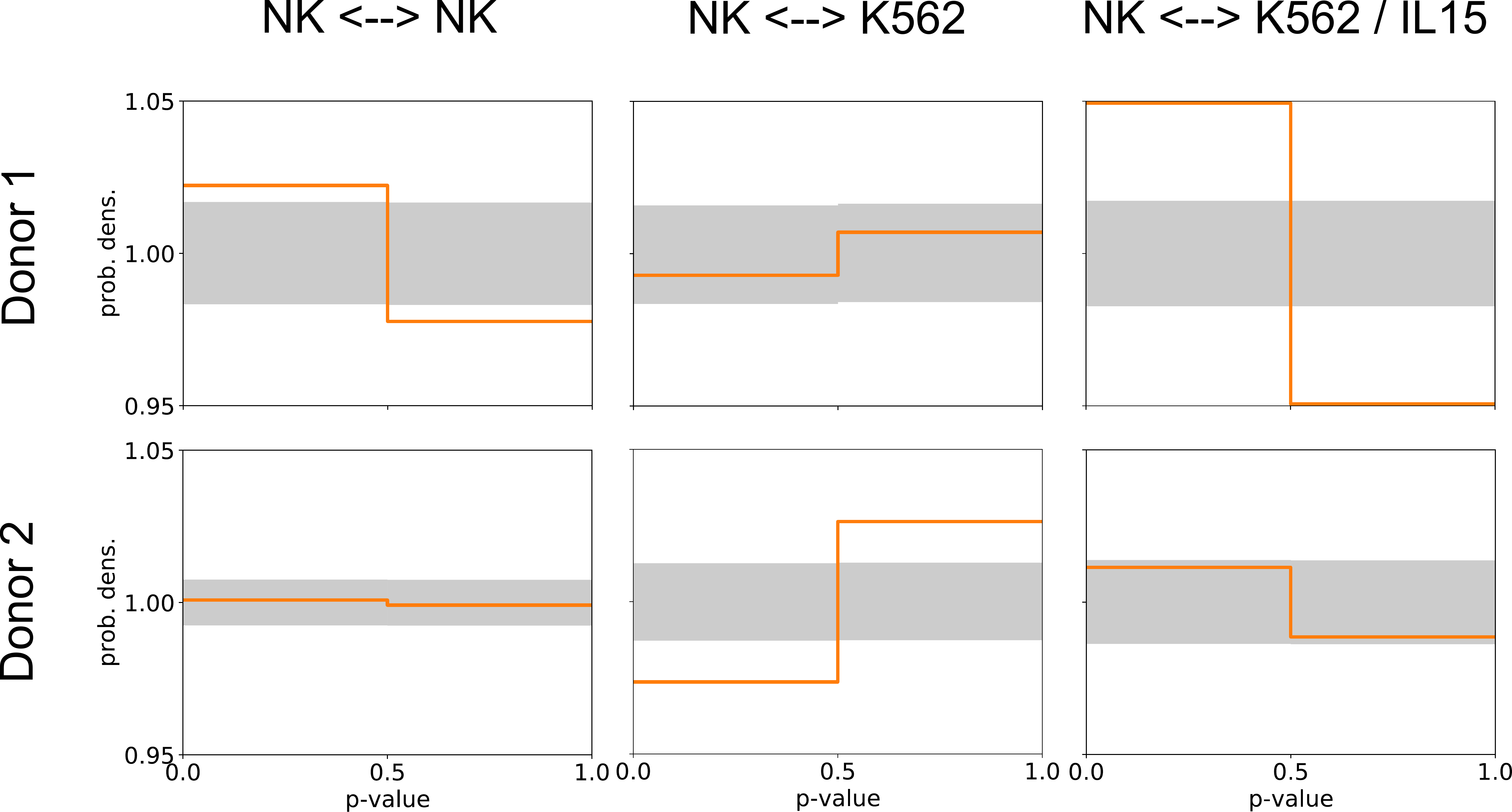}
\caption{p-value distributions of NK cells in a thick collagen gel. Shown are the interactions between NK cells and other NK cells (left column), between NK and wild-type K562 cells (center column), and between NK and K562 cells that secret the chemokine IL-15 (right column). Identical experiments have been performed using NK cells from two different donors (rows). To improve the statistics, the histograms have been computed with only two bins. For donor 1, we find (weak) attractive interactions between NK cells and strong attractive interactions between NK and IL-15-secreting K562 cells. For donor 2, we find repulsive interactions between NK and wild-type K562 cells.
\label{fig6}}
\end{figure}

Finally, we compute the p-value distributions of the NK cells for the three types of systems and for the two donors (Fig.~\ref{fig6}). For this evaluation, we use only trajectories with at least 20 consequtive entries, a maximum interaction radius of 100 $\mu$m, and 300 random reference distributions to compute the confidence intervals. Since the total amount of data is relatively small in these experiments, we limit the number of histogram bins to two, in order to improve the statistics in each bin. 
In the case of donor 1, we find attractive interactions between the NK cells, no significant interactions between NK and wild type K562 cells, but attractive interactions between NK and IL15-secreting K562 cells. By contrast, donor 2 shows no NK-NK interactions, repulsive interactions between NK and wild-type K562 cells, and no interactions between NK and IL15-secreting K562 cells (In the latter case, however, the measured probabilities of the two bins are very close to the borders of the confidence intervals, thus providing a weak evidence for attractive interactions).


\section*{Discussion}

Groups of animals, and even colonies of micro-organisms, often show interesting patterns of collective motion that can be theoretically replicated in the framework of self-driven, interacting agents \cite{vicsek1995novel}, provided the distance-dependent rules of interaction between the agents are known. For this purpose, various methods have been developed to extract the rules of interaction directly from motion data \cite{ballerini2008interaction,
lukeman2010inferring,
katz2011inferring,
eriksson2010determining,
mann2011bayesian,
gautrais2012deciphering,
attanasi2014information}, but (to our knowledge) it has not yet been attempted to extract long-range interactions between different cell types with a predator-prey relation. We have therefore tested different approaches to detect and quantify remote interactions between immune and tumor cells, based solidly on recorded cell trajectories. 

\vspace{0.2cm}
\noindent A first possible approach, used in some of the above studies \cite{eriksson2010determining,
mann2011bayesian}, is to set up an explicit model for the migration and interaction of the agents, and then to fit the unknown model parameters directly to the measured trajectories, for example using maximum likelihood optimization. We have applied such a method of parameter inference to simulated data of immune/tumor cell systems \cite{metzner2019inferring} and could correctly reproduce the known model parameters in some of the test cases. However, the inference produced wrong results whenever the investigated system had properties not fully captured by the assumed migration and interaction model - unfortunately a common situation in biology.

\vspace{0.2cm}
\noindent In this paper, we have therefore developed a new method of interaction detection based on p-values, which does not presume any detailed model of cell behavior, but only assumes that target-directed cells reveal themselves by a larger fraction of extraordinary, target-directed turns. Our method has only two user-adjusted parameters which slightly affect the results, namely the recording time interval $\Delta t$ and the maximum expected interaction range $r_{max}$. We have demonstrated that this p-value method reliably distinguishes between target-blind migration and purposeful pursuit in all test cases investigated so far. 

\vspace{0.2cm}
\noindent Recently, the misuse of p-values has been strongly criticized in the scientific community \cite{schmidt1997eight, gelman2006difference, goodman2008dirty, johnson2013revised, kyriacou2016enduring}. Indeed, many research studies consider the p-value as a fixed number that is uniquely determined by the experimental setup, although it actually is a random number drawn from a probability distribution \cite{sackrowitz1999p, taleb2016short}: When the very same experiment is repeated (that is, when new samples are drawn from the very same statistical model), the p-value will fall sometimes below and sometimes above the significance level. Picking just a single p-value thereby leads to non-reproducible results. For this reason, we compute the probability distribution $q_{obs}(\hat{p})$ of p-values (approximated by an empirical histogram with bins of a given size), pooling over all recorded steps of the immune cells. We then compare this observed distribution with the distribution $q_{ref}(\hat{p})$ of randomized reference systems. In order to estimate the statistical fluctuation of the reference distribution, we compute $q_{ref}(\hat{p})$ for a large ensemble of random reference systems, thus yielding a confidence interval for each bin of the p-value histogram. If there are interactions between immune and target cells, the observed p-values of the immune cells will be found outside of the confidence interval in at least some histogram bins. 

\vspace{0.2cm}
\noindent As a practical test, we have applied our p-value method to actual cells migrating in collagen matrices. In an initial experiment, human-derived NK cells were mixed with wild-type K562 cells in thin collagen gels that could be treated as effectively two-dimensional (see Supplemental Information). However, no evidence for long-range interactions was found in any of these eight initial data sets. We have therefore performed another experiment in thick, three-dimensional collagen gels, this time comparing a pure NK-system, a mixed NK-K562 system (wild-type tumor cells), and a mixed NK-K652/IL15 system (IL-15-secreting tumor cells). These three systems were measured with NK cells from two different human donors, and the p-value method revealed long-range interactions in three of the six resulting cases: Donor 1 showed attractive interactions between the NK cells, as well as attractive interactions between NK and IL-15-secreting K562 cells, whereas donor 2 showed repulsive interactions between NK and wild-type K562 cells.

\vspace{0.2cm}
\noindent It cannot be ruled out at this point that the contrasting outcomes between the first and second experiment are due to a different behavior of the cells in thin and thick collagen gels. However, considering also the contrasting results for the two donors in the second experiment, the non-uniform outcomes across all our measurements are most likely due to variability in the behavior of NK cells among human donors. Although this hypothesis still needs to be substantiated with further experiments, it would be of considerable relevance for immunotherapies if the chemotactic ability of immune cells varies drastically between humans and if this ability could be quantified using the p-value method.

\vspace{0.2cm}
\noindent The pattern of interactions found for donor 1 points to a scenario where NK immune cells are unable to sense and approach K562 tumor cells from afar, unless the tumor cells are manipulated to emit a traceable chemo-attractant, such as IL-15. However, once NK cells have established steric contact with tumor cells (by chance or with the help of a chemo-attractant), it may be necessary to summon further immune cells to the position of these targets, and the attractive interactions found between the NK cells of donor 1 provide just the required mechanism for this 'recruitment' process. By contrast, the repulsive interactions between NK and wild-type K562 cells in donor 2 are counter-intuitive at first glance. However, even such a behavior could be beneficial for the functioning of the immune system: Once a NK cell has found a tumor cell by chance and initiated its killing, there is no point in staying in that region any longer, and thus repulsive interactions that drive the NK cell away from the target directly after an attack would contribute to the overall efficiency of the immune system. In principle, this hypothesis could be tested by demonstrating that the repulsive interactions become only active for a short time after killing events. Detailed investigations of this kind might significantly extend our knowledge of NK cell behavior, but are beyond the scope of the present paper.

\vspace{0.2cm}
\noindent In summary, we have provided a new method for interaction-detection in systems with different cell types. The method was validated with simulated trajectories of chemotactic agents, and a subsequent application to experimental data has revealed first evidence for attractive and repulsive interactions. We hope that our preliminary results will stimulate further experiments and quantitative evaluations based on the p-value method, which might eventually contribute to a better understanding of the immune system and the development of related therapies.



\section*{Additional information}

\noindent{\bf Author contributions statement:}
CMe has devised the study, developed and implemented the methods for detecting interactions, applied the methods to the data, and wrote the paper. FHö and TC have performed the measurements. CMa has developed and implemented a cell tracking algorithm based on local entropy differences. AWi has trained a neural network for cell tracking. CVo has extracted, in vitro activated and expanded the immune cells. BFa has developed the imaging system and supervised the generation of the raw data. All authors reviewed the manuscript. \vspace{0.5cm}

\noindent{\bf Funding:}
This work was funded by the Grant ME1260/11-1 (347962689) of the German Research Foundation DFG.  \vspace{0.5cm}

\noindent{\bf Competing interests statement:}
The authors declare no competing interests.  \vspace{0.5cm}

\noindent{\bf Data availability statement:}
All data and the Python implementation of the p-value method are available online at \url{https://tinyurl.com/cm-pvaluemethod}.
\vspace{0.5cm}

\noindent{\bf Ethical approval and informed consent:}
See methods.  \vspace{0.5cm}

\noindent{\bf Third party rights:}
Data used in the paper and developed programs for cell imaging, tracking, and for computing the p-value distributions are the intellectual property of the authors.  \vspace{0.5cm}

\bibliographystyle{unsrt}
\bibliography{references}


\clearpage

\clearpage


\clearpage


\clearpage


\clearpage


\clearpage


\clearpage



\clearpage
\pagestyle{empty}

\begin{center}
{\huge \bf Supplemental Information}\\ \vspace{1.5cm}
{\huge Detecting long-range interactions \\between migrating cells} \\ \vspace{1cm}
{\large 
Claus Metzner$^{*}$, 
Franziska Hörsch$^{*}$,
Christoph Mark$^{*}$,\\
Tina Czerwinski$^{*}$,
Alexander Winterl$^{*}$,\\
Caroline Voskens$^{+}$,
and Ben Fabry$^{*}$} \\ \vspace{0.5cm}
{$^{*}$Biophysics, Friedrich-Alexander University Erlangen-Nürnberg}\\ 
{$^{+}$Dermatology, University Hospital Erlangen}\\ \vspace{1cm}
{Correspondence to {\em claus.metzner@gmail.com}}\\
\end{center}


\clearpage
\subsection*{Data measured with NK and K562 cells in flat collagen gels}

\begin{figure}[h!]
\centering
\includegraphics[width=15cm]{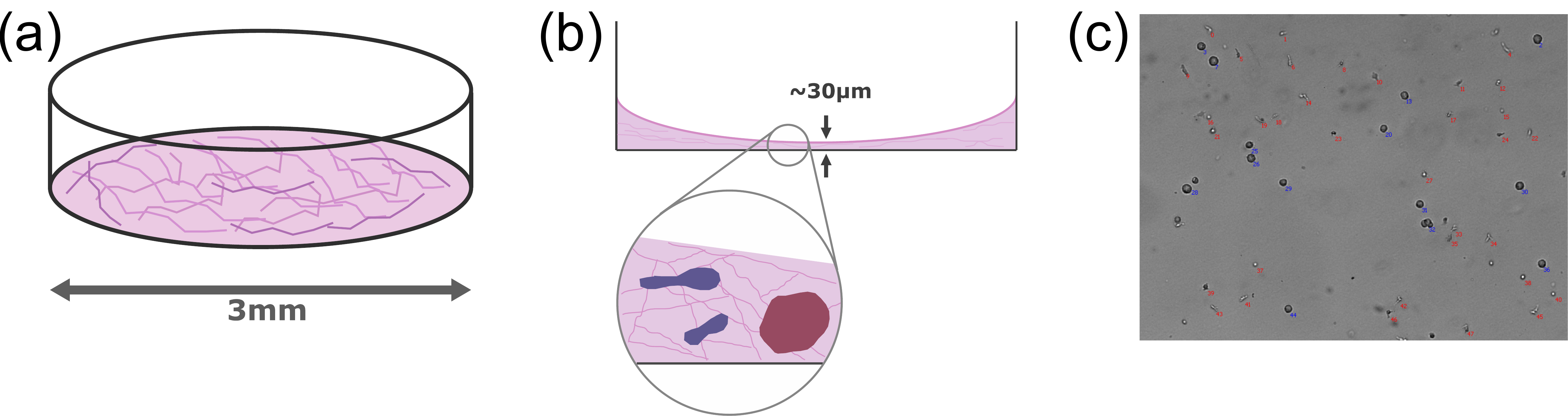}
\caption{(a,b) Experimental setup for flat gels. (c) Example frame with NK cells in red and K562 cells in blue.
\label{fig7}}
\end{figure}

In addition to the data presented in the main part of the paper, we have analyzed 9 additional data sets DS0-DS8 (see table \ref{DS_list}). The two human donors of the NK cells used in these 9 data sets were different from those in the main part of the paper. The experiments were this time performed in thin, quasi-two-dimensional matrices, and also the cell tracking was performed in a different way:

\vspace{0.5cm}
\noindent In data sets DS0-DS8, the NK cells were again in vitro activated and expanded. A number of $5\cdot 10^6$ NK immune cells and $3\cdot 10^6$ K562 tumor cells are mixed with ice-cold 500 $\mu$l acid-dissolved collagen solution (1.2mg/ml) and pipetted into a tissue-culture-treated 35 mm dish (Fig.~\ref{fig7}(a); for a detailed protocol, see \cite{Metzner2015}). The polymerization of the collagen solution is initiated by placing the dish for 30 min in a cell culture incubator at 37$^\circ$C, 5$\%$ CO\textsubscript2. Due to surface tension, the thickness of the polymerized collagen gel decreases towards the center of the dish with a height of $\approx$ 30 $\mu$m (Fig.~\ref{fig7}(b)). Time-lapse imaging can thus be realized in bright-field mode without scanning in z-direction, while the cells still showed the same characteristic migration behavior as in a thick collagen gel. We recorded 9 independent data sets, each including between 333 and 1547 images, with a time interval of 45 seconds between two subsequent frames. The images had 1344 x 1024 pixels with a linear size of 0.645 $\mu$m. 

\vspace{0.5cm}
\noindent In data sets DS0-DS8, cells are automatically segmented using local differences of image entropy. The classification in immune and tumor cells is based on differences in speed, size, and brightness (Fig.~\ref{fig7}(c)). Finally, the temporal trajectory of each cell is determined using the overlap of the cell area between successive time frames.

\vspace{0.5cm}
\begin{table}[h!]
\begin{center}
\begin{tabular}{| l | r | r | r | r | r | r | r |}\hline
DS & $N_{fra}$ & $T_{rec}$ (h) & $N_{tri}$ & $N^{(0)}_{imm}$ & $N^{(0)}_{tum}$ & $\overline{v}=\frac{\overline{\sigma}}{\Delta t}$ ($\frac{\mu m}{min}$) & $\overline{\kappa}$\\ \hline
0 & 334 & 4.18 & 12075 & 34 & 14 & 5.96 & 2.59 \\ \hline
1 & 414 & 5.18 & 6910 & 23 & 27 & 4.87 & 2.71 \\ \hline
2 & 641 & 8.01 & 11166 & 22 & 17 & 6.75 & 2.37 \\ \hline
3 &1548 &19.35 & 58736 & 34 & 14 & 5.07 & 2.57 \\ \hline
4 & 633 & 7.91 & 29101 & 47 & 18 & 4.12 & 1.73 \\ \hline
5 & 640 & 8.00 & 24272 & 26 & 26 & 6.39 & 3.28 \\ \hline
6 & 640 & 8.00 & 10561 & 16 & 20 & 5.29 & 2.72 \\ \hline
7 & 640 & 8.00 & 12590 & 19 & 23 & 6.20 & 2.67 \\ \hline
8 & 640 & 8.00 & 8095 & 13 & 23 & 4.92 & 2.59 \\ \hline
\end{tabular}
\caption{
\label{DS_list}  
Essential properties of the nine data sets $DS=0\ldots8$. Here, $N_{fra}$ is the number of video frames, $T_{rec}$ the total recording time in hours, $N_{tri}$ the number of valid triplets that could be used for the p-value evaluation, $N^{(0)}_{imm}$ the initial number of immune cells, $N^{(0)}_{tum}$ the initial number of tumor cells, $\overline{v}$ the average speed of immune cells in $\mu$m/min, and $\overline{\kappa}$ is the average persistence parameter of immune cells.
}
\end{center}
\end{table}

\clearpage
\subsubsection*{p-value distributions of NK cells with wildtype K562 cells (data sets DS0-DS8)}

\begin{figure}[h!]
\centering
\includegraphics[width=17cm]{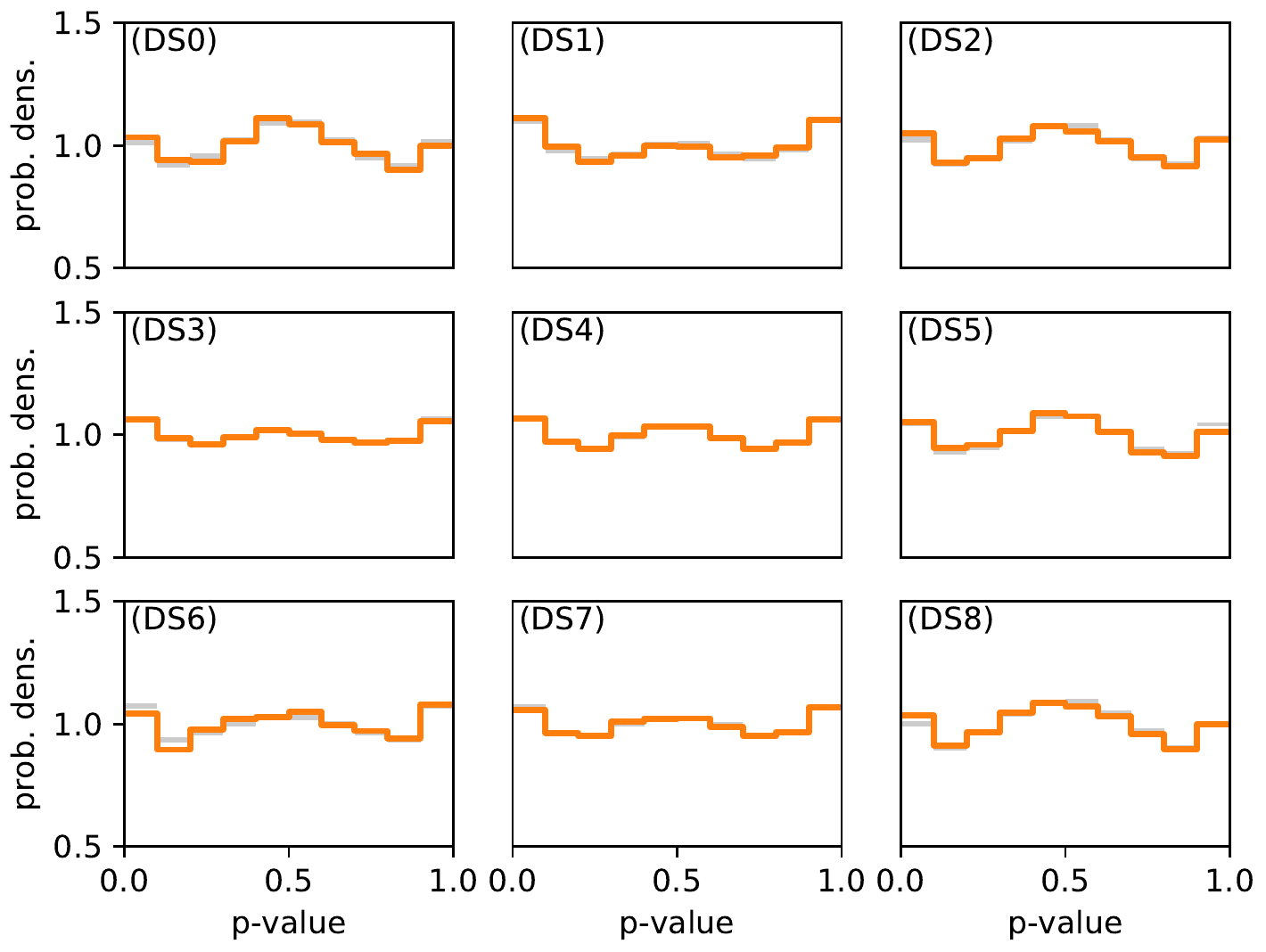}
\caption{p-value distributions for data sets DS0-DS8, using a recording time interval of $\Delta t = 0.75$min and a maximum interaction radius of $r_{max}=700\mu$m.
\label{fig8}}
\end{figure}

We do not find evidence for interactions between NK cells and wild-type K562 cells in any of the data sets DS0-DS8 (Fig.~\ref{fig8}).

\clearpage
\subsubsection*{Effect of recording interval and max. interaction distance (data set DS0)}

\begin{figure}[h!]
\centering
\includegraphics[width=17cm]{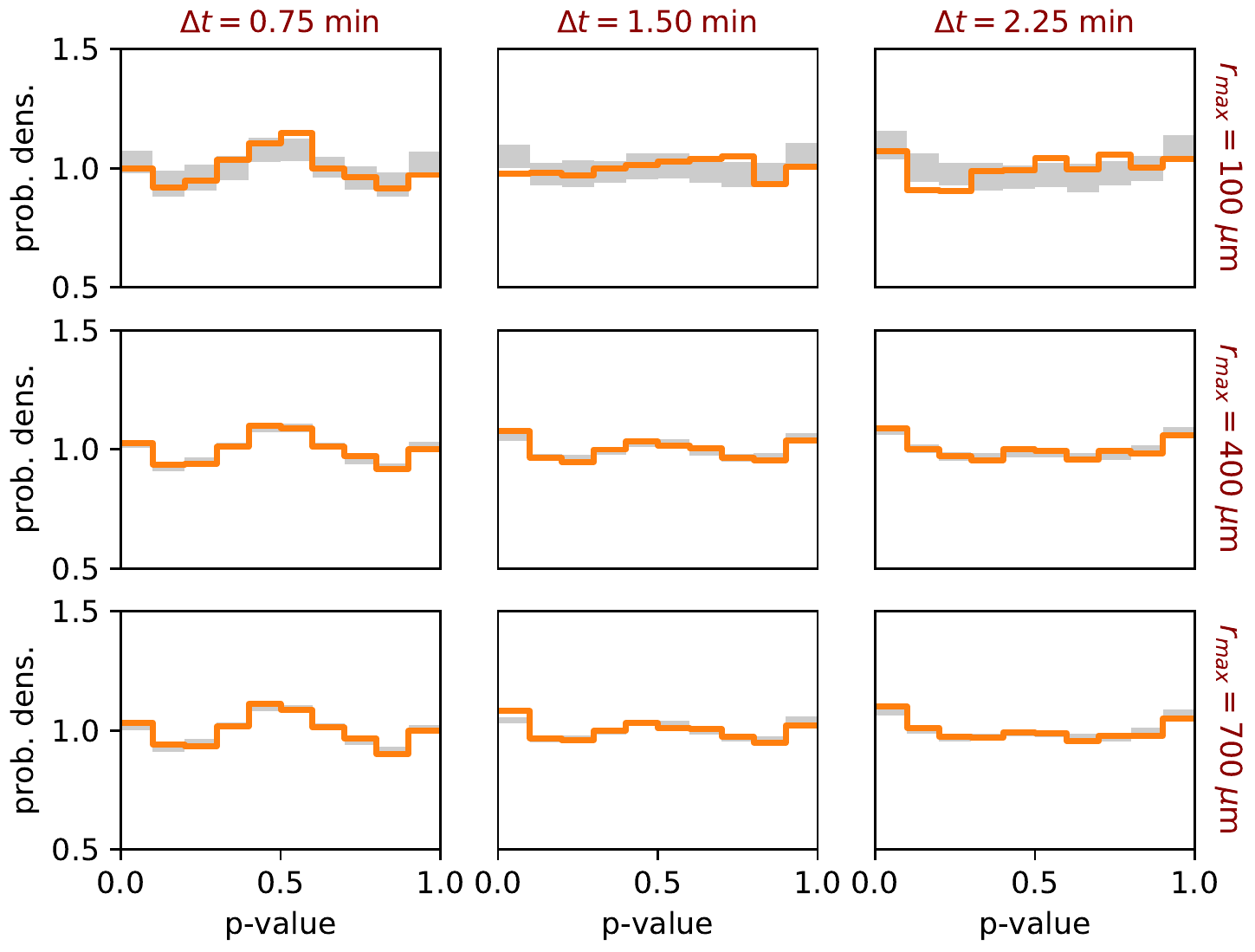}
\caption{p-value distributions of NK cells with wild-type K562 cells (data set DS0). 
\label{fig9}}
\end{figure}

We do not find evidence for interactions between NK cells and wild-type K562 cells, independent of the recording interval and the maximum interaction distance
(Fig.~\ref{fig9}).

\clearpage
\subsubsection*{p-value distributions for NK self-interaction (data sets DS0-DS8)}

\begin{figure}[h!]
\centering
\includegraphics[width=17cm]{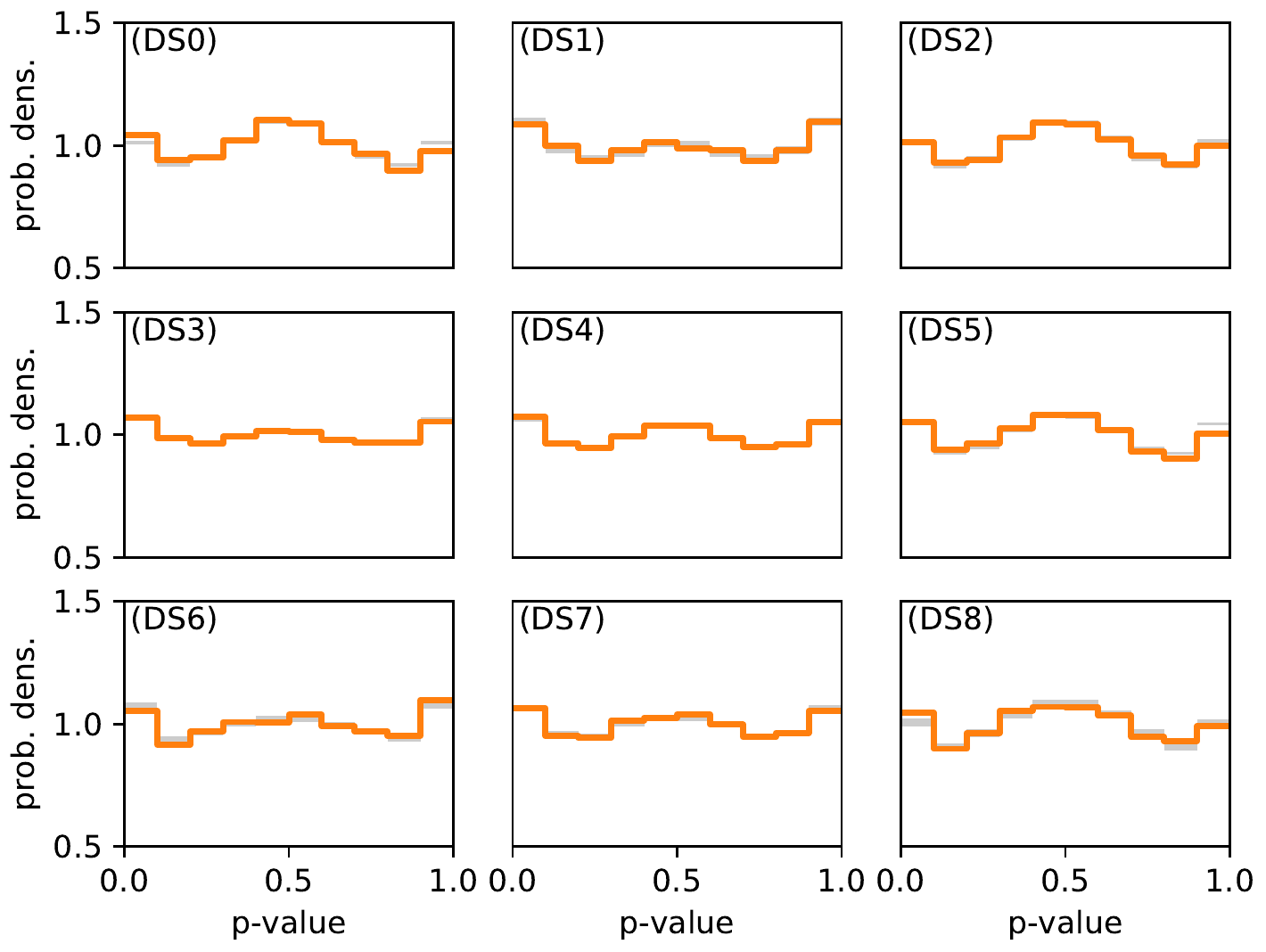}
\caption{p-value distributions for the interactions of NK cells among themselves (data sets DS0-DS8), using a recording time interval of $\Delta t = 0.75$min and a maximum interaction radius of $r_{max}=700\mu$m.
\label{fig10}}
\end{figure}

Our p-value method can also be applied to systems with more than two different cell types, and it is not necessary to know the 'predator-prey relations' in advance. To demonstrate this feature, we have used data sets DS0-DS8 to investigate possible interactions between the NK cells themselves. However, there is no clear evidence for such interactions (Fig.~\ref{fig10}).

\clearpage
\subsubsection*{Video material}

To compare the experimentally observed cell behavior with the models of blind search and spatial gradient sensing,
we provide three videos (\url{https://tinyurl.com/cm-pvaluemethod}).

\vspace{0.2cm}
\noindent The video {\bf V1.mp4} shows the tracked cells of our data set DS0. The NK immune cells are shown as red circles, the K562 tumor cells as blue circles. All cells are labeled with unique numbers. Once a tumor cell is visited by an immune cell (within a distance smaller then $30\; \mu m$), the tumor cell is considered as 'found' and is subsequently colored in gray.

\vspace{0.2cm}
\noindent The video {\bf V2.mp4} shows a simulation that starts with the same initial configuration as in data set 0. The simulated immune cells also migrate with the same average speed and directional persistence as in the experiment (Tumor cells are assumed to by completely stationary for simplicity). This simulation assumes a target-blind random walk (blind search BLS) and produces a rate of encounters between immune and tumor cells comparable to that in data set 0.

\vspace{0.2cm}
\noindent The video {\bf V3.mp4} shows a simulation analogous to V2.mp4, however assuming that the immune cells are chemotactically active and sense the tumor cells using spatial gradient sensing (SGS). This systematic way of approaching the targets leads to a significantly higher rate of encounters. 



\end{document}